\documentclass[12pt]{article}
\usepackage[dvipdfmx]{graphicx}
\usepackage{amsfonts,amsmath,amssymb,epsf,cite}
\usepackage{amsmath,braket}
\usepackage{graphicx,color}
\newcommand{\de}{\partial}
\newcommand{\be}{\begin{equation}}
\newcommand{\ba}{\begin{eqnarray}}
\newcommand{\ea}{\end{eqnarray}}
\newcommand{\ee}{\end{equation}}

\newcommand{\f}{\frac}
\newcommand{\s}{\sqrt}

\newcommand{\ap}{\alpha}

\newcommand{\ddd}{\cdot\cdot\cdot}
\newcommand{\no}{\nonumber \\}
\newcommand{\la}{\langle}
\newcommand{\lb}{\rangle}
\newcommand{\bea}{\begin{eqnarray}}
\newcommand{\eea}{\end{eqnarray}}
\newcommand{\bes}{\begin{equation*}}
\newcommand{\beas}{\begin{eqnarray*}}
\newcommand{\eeas}{\end{eqnarray*}}
\newcommand{\bas}{\begin{array*}}
\newcommand{\eas}{\end{array*}}
\newcommand{\ees}{\end{equation*}}

\newcommand{\ep}{\epsilon}

\begin{document}

\begin{titlepage}
\thispagestyle{empty}

\begin{flushright}
YITP-19-85
\\
IPMU19-0123
\\
\end{flushright}


\begin{center}
\noindent{{\textbf{Universal Local Operator Quenches and Entanglement Entropy}}}\\
\vspace{2cm}
Arpan Bhattacharyya$^{a}$, Tadashi Takayanagi$^{a,b}$ and Koji Umemoto$^{a}$
\vspace{5mm}

{\it
$^{a}$ Center for Gravitational Physics, Yukawa Institute for Theoretical Physics,\\
Kyoto University, Kyoto 606-8502, Japan\\
$^{b}$ Kavli Institute for the Physics and Mathematics of the Universe,\\
University of Tokyo, Kashiwa, Chiba 277-8582, Japan\\
}

\vskip 2em
\end{center}

\begin{abstract}
We present a new class of local quenches described by mixed states,
parameterized universally by two parameters.
We compute the evolutions of entanglement entropy for both a holographic and Dirac fermion
CFT in two dimensions. This turns out to be equivalent to calculations of two point functions on a torus. 
We find that in holographic CFTs, the results coincide with the known results of pure state 
local operator quenches. On the other hand, we obtain new behaviors
in the Dirac fermion CFT, which are missing in the pure state counterpart. By combining our results 
with the inequalities known for von-Neumann entropy, we obtain an upper bound of
the pure state local operator quenches in the Dirac fermion CFT. We also explore predictions about
the behaviors of entanglement entropy for more general mixed states.
\end{abstract}

\end{titlepage}

\newpage

\tableofcontents

\newpage

\section{Introduction}

Entanglement entropy characterizes the degrees of freedom in quantum field theories (QFTs) and thus plays a role of the universal order parameter in QFTs \cite{CCR,CH,Ni}.  At the same time, entanglement entropy is an important probe of spacetime geometry via the AdS/CFT \cite{RT,HRT}. Properties of entanglement entropy in quantum field theories have been extensively studied so far. However, in most of the works so far, each of total systems is given by a pure state given by a vacuum or a
certain excited state. Examples for which total systems are given by mixed states have been
mostly limited to finite temperature states. The main purpose of this paper is to introduce an interesting new class of mixed states in conformal field theories (CFTs) and present calculations of entanglement entropy for such mixed states. Consider the von-Neumann entropy of a mixed state $\rho$, which is written as a linear combination:
\be
\rho=\sum_i p_i\rho_i,
\ee
where $\rho_i$ $(i=1,\ddd,N)$ are density matrices and the coefficients $0<p_i\le1$ satisfy
$\sum_i p_i=1$. The concavity inequality and another well-known inequality give the lower and upper bound of the entropy $S(\rho)$ as follows (refer to e.g. \cite{Book})
\ba
\sum_i p_i S(\rho_i)\leq S(\rho)\leq \sum_i p_i S(\rho_i)+H(p),  \label{eeb}
\ea
where $H(p)=-\sum_i p_i\log p_i$ is the Shannon entropy. The left inequality is saturated when
$\rho_i$ are all identical, while the right one is saturated when $\rho_i$ are orthogonal to each other.
The inequality (\ref{eeb}) gives an important clue to understand properties of entanglement entropy for mixed states. To see this, consider a holographic setup where two classical gravity backgrounds $M_1$ and $M_2$ are dual to two pure quantum states $\rho^{(1)}=|\psi_1\lb \la \psi_1|$ and $\rho^{(2)}=|\psi_2\lb \la \psi_2|$, respectively. Let us ask what we can say about the mixed state $p\rho^{(1)}+(1-p)\rho^{(2)}$. To study this, let us consider a reduced density matrix $\rho_A=p\rho^{(1)}_A+(1-p)\rho^{(2)}_A$ for a subsystem $A$. Since the Shannon entropy $H(p)$ is $O(1)$ and $S(\rho_i)$ is the order of the central charge $c$ (or equally the square of the rank of gauge group $N^2$), the inequality (\ref{eeb}) approximately becomes an equality at the leading order of $1/c=1/N^2$ expansion:
\ba
S(\rho_A)\simeq pS(\rho^{(1)}_A)+(1-p)S(\rho^{(2)}_A).  \label{linearh}
\ea
The same is true for a linear combination of states as long as the number of states is much smaller than $c$.
This property in holographic CFTs (\ref{linearh}) can be interpreted as the linearity of area operator in
the AdS/CFT, which has been studied for explicit examples in \cite{ADSw}.

However, when we take a linear combination of many states whose total numbers are $O(c)$,
we cannot neglect the Shannon term and the inequality does not become tight.
This is the case where we are interested in this paper.

In particular, we will choose $\rho_i$ to be a locally excited state in a CFT, created by acting a local operator $O_i(x)$ on a vacuum. This requires an appropriate regularization, which introduces an infinitesimally small parameter $s$. One interesting class of examples is provided by choosing $p_i$ to be Boltzmann-like distribution $p_i\propto e^{-\beta \Delta_i}$, where $\Delta_i$ is the conformal dimension of the operator $O_i(x)$. Note that this is different from the genuine Boltzmann distribution which describes a finite temperature state because $\Delta_i$ is not the energy of the locally excited state on R$^2$ (i.e. a CFT on an infinite line). This defines a very simple but new class of (time-dependent) mixed states in CFTs in any dimensions.

In this paper we will focus on two dimensional CFTs on a plane R$^2$ and analyze the properties of this class of mixed states, especially through computations of the entanglement entropy of $\rho_A=\sum p_i\rho^{(i)}_A$. This class of mixed states only depend on two parameters: $\beta$ and the regularization parameter $s$ of the local operator. Note that implicitly the result depends on the standard UV cut off (or lattice constant) denoted by $\ep$ as usual in the calculations of 
entanglement entropy. 

Even though the properties of each of locally excited states $\rho_i$ depend on the details of $O_i(x)$,
the mixed states we consider only depend on the two parameters $\beta$ and $s$ as we mentioned. Motivated by this, we call this class of mixed states universal local operator quenches. We expect that properties of universal local quench states, e.g. the time evolutions of entanglement entropy, may provide a classification of CFTs, which can distinguish chaotic properties between different CFTs. This will be confirmed in the results of this paper.

This paper is organized as follows: In section 2, we present a brief review of local operator quenches based on each pure state. In section 3, we introduce the universal local operator quenches based on mixed states and describe the path-integral description as well as computations of entanglement entropy.
In section 4, we study the time evolution of entanglement entropy under the universal local quenches in 2d holographic CFTs. In section 5,  we study the time evolution of entanglement entropy under the universal local quenches in the 2d Dirac Fermion CFT. In section 6, we analyze the consequence of the bounds (\ref{eeb}) for our results of  universal local quenches. In section 7, we discuss similar bounds for R\'{e}nyi entropy and presents examples of explicit calculations of R\'{e}nyi entropy for simple mixed states with local excitations. In section 8, we explore the entanglement entropy for more general mixed states based on the inequality (\ref{eeb}). In section 9 we summarize conclusions and discuss future
problems.

\section{Review of Pure State Local Operator Quenches}

An instantaneous excitation at a point is called a local quench.
A fundamental and important class of local quenches is the local operator quenches \cite{Nozaki},
which is created by acting a local operator $O_i(x)$ on the vacuum (this pure state is denoted by $\rho_i$):
\ba
|\Psi_i(t)\lb={\cal{N}}_i e^{-itH}e^{-sH}O_i(x)|0\lb,  \label{lexc}
\ea
where $H$ is the CFT Hamiltonian.\footnote{There are other types of local quenches such as the joining local quenches 
\cite{CCL} and the splitting local quenches \cite{Shimaji:2018czt}. The holographic dual of joining quenches is given by 
\cite{Ugajin}.}
 We write the conformal dimension of the operator $O_i(x)$
as $\Delta_i(=h_i+\bar{h_i})$, where $h_i$ and $\bar{h}_i$ are its chiral and anti-chiral conformal dimension.
We chose $\{O_i(x)\}$ to be an orthonormal basis such that
\be
\la O_i(x)O_j(y)\lb=\delta_{ij}\left(\frac{\ep}{x-y}\right)^{2\Delta_i},
\ee
where $\ep$ is the standard UV cut off (or lattice constant) denoted by $\ep$.
The parameter $s$ in (\ref{lexc}) is infinitesimally small and provides the regularization
for the local quench procedure.\footnote{This local quench regularization parameter $s$ should be distinguished from the ordinary UV cut off $\ep$, which is the lattice spacing. For example, the ground state entanglement entropy when the subsystem $A$ is an interval with the length $L$ is given by the well-known result $S^{(0)}_A=\frac{c}{3}\log \frac{L}{\ep}$.} The normalization factor ${\cal{N}}_i$ is chosen such that $\la\Psi_i(t)|\Psi_i(t)\lb=1$, which leads to
${\cal{N}}_i=\left(\frac{2s}{\ep}\right)^{\Delta_i}$.

Here we are interested in the entanglement entropy for the locally excited state (\ref{lexc}) for the subsystem $A$ defined by an interval $[a,b]$. The complement of $A$ is written as $B$. The entanglement entropy
depends on the time $t$ as well as the regularization parameters $s$ and $\ep$. The entanglement entropy is computed as
\ba
S_A=S(\rho^{(i)}_A),
\ea
where $\rho_i$ is defined as
\be
\rho^{(i)}_A=\mbox{Tr}_B\left[|\Psi_i(t)\lb\la\Psi_i(t)|\right].  \label{lqop}
\ee

We can calculate the entanglement entropy $S(\rho^{(i)}_A)$ via
the standard replica method:
\ba
S(\rho^{(i)}_A)&=&-\frac{\de}{\de n}\log\left[\mbox{Tr}_A (\rho^{(i)}_A)^n \right]\bigr|_{n=1}\no
&=&-\frac{\de}{\de n}\log\left[\frac{\la O_i(is+t)O_i(-is+t)\sigma_{n}(a)\bar{\sigma}_{n}(b)\lb}{\la O_i(is+t)O_i(-is+t)\lb}\right] \Biggr|_{n=1}, \label{replicaa}
\ea
where $\sigma_n$ denotes the twist operator. In particular, we take the large subsystem size limit and focus on the following time region:
\ba
0<s\ll a\ll t\ll b.   \label{middle}
\ea
In the time region $a<t<b$, the entanglement entropy gets non-trivial because
 the excitations created by the local operator propagates into both the subsystem $A$ and $B$ at the same time, 
due to the left and right-moving modes at the speed of light.

In the papers \cite{Nozaki,HNTW},  this entropy was computed for the primary states for free CFTs and rational CFTs
 and it was found that the entropy is increased only by a finite amount,
which coincides with the quantum dimension in 2d CFTs.

For holographic 2d CFTs,  the entropy was computed in \cite{NNT} using AdS/CFT, which was reproduced in \cite{Hat} from the analysis for large $c$ CFT. Assuming that $\Delta_i$ is $O(c)$,
this result in holographic CFTs reads
\ba \label{entexp}
S(\rho^{(i)}_A)\simeq \frac{c}{3}\log\frac{b-a}{\ep}+\frac{c}{6}\log \frac{t}{s}
+\frac{c}{6}\log\left[\frac{1}{\s{\frac{12\Delta_i}{c}-1}}\sinh\left(\pi\s{\frac{12\Delta_i}{c}-1}\right)\right].
\label{eeploq}
\ea
If $\Delta_i$ is much smaller than the central charge $c$, then the above formula gets modified. 
In particular if we take the limit $\Delta_i\to 0$, $S(\rho^{(i)}_A)$ coincides with the entanglement entropy for the ground state \cite{HLW} 
\ba
S^{(0)}_A=\frac{c}{3}\log\frac{b-a}{\ep}.  \label{vacentanglement entropy}
\ea

Note that the above result (\ref{eeploq}) in holographic CFTs is universal. The entanglement entropy only depends on the conformal dimension and is independent of the other details of the local operators. This behavior is peculiar
to holographic CFTs, while the entanglement entropy  in general
CFTs depends on various details of local operators.\footnote{For a partial list of progress on properties of entanglement entropy under local operator quenches refer to \cite{Shimaji:2018czt,CNT, Caputa:2014eta,deBoer:2014sna,Guo:2015uwa,Chen:2015usa,Nozaki:2015mca,Caputa:2015tua,Caputa:2015waa,Rangamani:2015agy,
Sivaramakrishnan:2016qdv,Caputa:2016yzn,Numasawa:2016kmo,Nozaki:2016mcy,David:2016pzn,Caputa:2017tju,Nozaki:2017hby,JaTa,He:2017lrg,KuTa,Ku,Apolo:2018oqv,Kusuki:2019gjs,Caputa:2019avh,He:2019vzf,Kusuki:2019avm}.}

\section{Universal Local Operator Quenches}

Consider the following mixed state $\rho(\beta,s)$ as a local operator quench state:
\be
\rho(\beta,s)=\sum_i  \frac{e^{-\beta \left(\Delta_i-\frac{c}{12}\right)}}{Z(\beta)}\cdot |\Psi_i(t)\lb\la\Psi_i(t)|,  \label{mixL}
\ee
where
\be
Z(\beta)=\sum_{i}e^{-\beta\left(\Delta_i-\frac{c}{12}\right)}.
\ee
The indices $i$ run over all the quantum states in a given CFT including both primary and descendant states. The state $ |\Psi_i(t)\lb$ is given by (\ref{lexc}) with the regularization parameter $s$.

Notice that this state is universally defined up to two parameters $\beta$ and $s$ for any
CFTs in any dimensions. Therefore we call this a universal local operator quench.
It is also useful to note that $Z(\beta)$ is the thermal partition function on a cylinder.
Below we will study the behavior of entanglement entropy for this state in two dimensional CFTs.

\subsection{Path-integral Description}

\begin{figure}
  \centering
  \includegraphics[width=10cm]{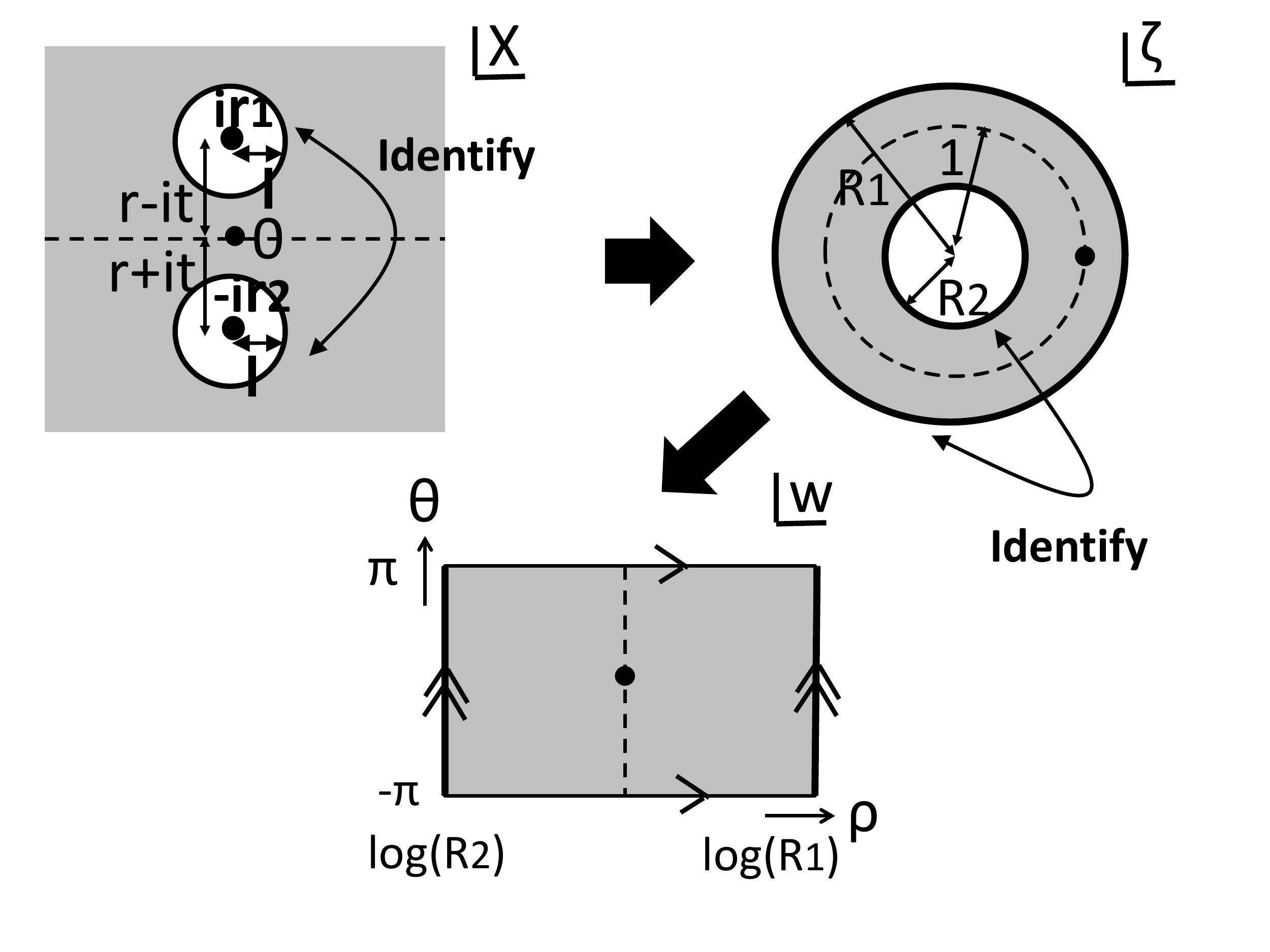}
  \caption{The path-integral on a torus $\Sigma$ which describes a mixed state local quench and its conformal transformations into 
$\zeta$ and $w$ coordinate from the original $X$ coordinate.}
\label{fig:twocirmap}
  \end{figure}

First we would like to provide a path-integral description of this mixed state $\rho(\beta,s)$ and the computations of the entanglement entropy $S_A=S(\rho_A)$, where $\rho_A=\mbox{Tr}_B[\rho(\beta,s)]$.
We introduce the complex coordinate $(X,\bar{X})$, where the universal local operator quench (\ref{mixL})
will take place.  We also write the coordinate as $X=x+i\tau$, where
$\tau$ is the Euclidean time. We pick up an annulus region, called $\Sigma'$
on this complex plane with two round disks (radius $l$) removed i.e. the region defined by
\be
\Sigma': \ \ |X-ir_1|\geq l,\ \ \ |X+ir_2|\geq l,  \label{antwc}
\ee
where $r_{1}$ and $r_2$ are given by
\be
r_1=r-it,\ \ \ r_2=r+it,  \label{rexpv}
\ee
where $r$ is a parameter related to the regularization parameter of local quench as we will clarify later,
and $t$ is the real time coordinate (we assume analytical continuation of the Euclidean time).
These are summarized in the upper left picture in Fig.\ref{fig:twocirmap}.

We find that the following conformal map from $X$ to $\zeta$:
\ba
&& X=i\left(\f{\ap_1\zeta+\ap_2}{\zeta+1}\right), \label{cmapa} \\
&&  \ap_1\equiv \f{r_1-r_2}{2}-\s{\f{(r_2+r_1)^2}{4}-l^2}, \no
&&  \ap_2\equiv \f{r_1-r_2}{2}+\s{\f{(r_2+r_1)^2}{4}-l^2}.
\ea
This transforms the two circles $|X-ir_1|=l$ and $|X+ir_2|=l$ into  two concentric circles centered at the origin
$|\zeta|=R_2$ and $|\zeta|=R_1$, where the two radii $R_1>R_2$ are
\ba
&& R_1=\f{l}{\f{r_1+r_2}{2}-\s{\f{(r_2+r_1)^2}{4}-l^2}},\no
&& R_2=\f{l}{\f{r_1+r_2}{2}+\s{\f{(r_2+r_1)^2}{4}-l^2}}. \label{radtim}
\ea

Using (\ref{rexpv}), we can rewrite this as follows:
\ba
&& \ap_1=-it-\s{r^2-l^2},\ \ \ \ap_2=-it+\s{r^2-l^2},\no
&& R_1=\f{r+\s{r^2-l^2}}{l}(>1),\ \ \ R_2=\f{r-\s{r^2-l^2}}{l}(<1).
\ea
Notice also the relation $R_1R_2=1$.

Now we glue the two boundary circles $|X-ir_1|= l$ and  $|X+ir_2|=l$ of
the annulus $\Sigma'$ (\ref{antwc}) together and
obtain a torus, denoted by $\Sigma$. 
A sketch of this torus $\Sigma$ in the coordinate $\zeta$ is depicted as the upper right picture
in Fig.\ref{fig:twocirmap}.

It is useful to further perform the standard transformation
\be
\zeta=e^{w}, \label{cmapb}
\ee
with $w=\rho+i\theta$. The annulus is now mapped into the torus whose coordinate takes
values in
\be
\log R_2 \leq \rho\leq \log R_1, \ \ 0\leq \theta \leq 2\pi, \label{onesi}
\ee
as in the lower picture in Fig.\ref{fig:twocirmap}.

\begin{figure}
  \centering
  \includegraphics[width=10cm]{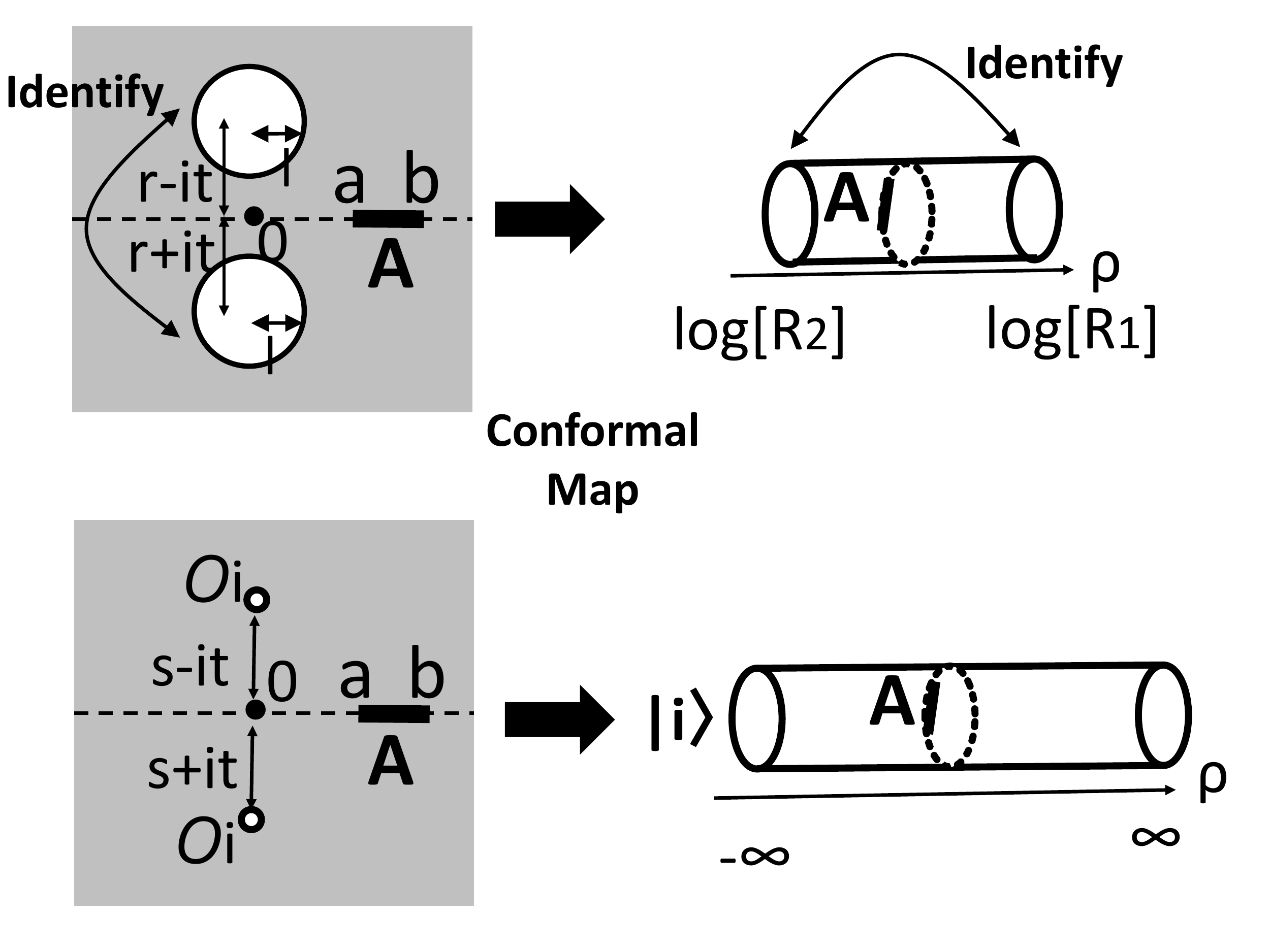}
  \caption{Comparison between the universal (mixed state) local quench  (\ref{mixL}), depicted in the upper picture 
and the pure state local quench (\ref{lexc}), depicted in the lower picture. We also
expressed the interval $A$ for which we calculate the entanglement entropy as the thick lines.}
\label{fig:comp}
  \end{figure}

We would like to argue that the universal local operator quench state $\rho(\beta,s)$ (\ref{mixL})
is described by a path-integral on this torus $\Sigma$.
Remember that the conformal transformations (\ref{cmapa}) and (\ref{cmapb})  map the region $\Sigma$ into a cylinder with the two end circles identified into a torus (refer to the upper pictures in Fig.\ref{fig:comp}). 
The length of the torus in the $\rho$ direction is
\be
\beta=\log\frac{R_1}{R_2}=\log\left[\frac{r+\s{r^2-l^2}}{r-\s{r^2-l^2}}\right],  \label{betaf}
\ee
which is identified with $\beta$ in (\ref{mixL}).
In the torus picture, we would like to regard $\rho$ as the Euclidean time. It is useful to notice that the original time slice $\tau=0$ corresponds to 
$\rho=0$.

Our claim that the path-integral on this torus $\Sigma$ describes the universal local quench state $\rho(\beta,s)$ can be understood by comparing this setup with the path-integral description of 
pure state counterpart. We can regard the pure state local quench state $|\Psi_i(t)\lb\la\Psi_i(t)|$ (\ref{lexc}) as the small hole limit $l\to 0$, with the identification $s=r$. As in lower pictures of Fig.\ref{fig:comp}, the pure state local quench is described by an infinitely long cylinder where the state $|i\lb$, dual to the operator $O_i$ via the usual state/operator correspondence, propagates.

To see more details, it is important to keep track of the lattice spacing.  We write the lattice spacing in the original coordinate $X$ as $\ep$. Note that $\ep$ takes the same value everywhere in this coordinate. This cut off is mapped into that in the torus coordinate $w$, which is denoted by $\delta$. As follows from our conformal transformations (\ref{cmapa}) and (\ref{cmapb}), we find that the cut off $\ep$ is related to $\delta$:
\be
\delta=\frac{2\s{r^2-l^2}}{\s{((x+t)^2+r^2-l^2)((x-t)^2+r^2-l^2)}}\cdot \ep. \label{uvm}
\ee
Similarly, the cut off relation for the pure state local quench reads
\be
\delta=\frac{2s}{\s{((x+t)^2+s^2)((x-t)^2+s^2)}}\cdot \ep.
\ee

Therefore, the cut off $\delta$ for the universal local operator quench state $\rho(\beta,s)$ (upper picture of Fig.\ref{fig:comp})
is related to that for the pure state one $|\Psi_i(t)\lb\la\Psi_i(t)|$  (lower picture of Fig.\ref{fig:comp}) via the identification of the
quench parameter
\be
s=\s{r^2-l^2}.   \label{rel}
\ee

In summary, after the conformal maps, the state for the path-integration over the torus $\Sigma$ can be identified with the state
$\rho(\beta,s)$ defined by  (\ref{mixL}), where the parameters $\beta$ and $s$
 are related to $r$ and $l$ in the path-integral description by (\ref{betaf}) and (\ref{rel}), respectively.

\subsection{Calculating Entanglement Entropy under Universal Local Operator Quench}

Now we move back to the original problem of computing entanglement entropy for the universal local operator quench states. We choose the subsystem $A$ to be an interval $x\in [a,b]$ on the real axis $\tau=0$, which is mapped into an interval on the circle
$\rho=0$. The UV cut off   (\ref{uvm}) in the torus coordinate which corresponds to the original one $\ep$ is denoted by $\delta_a$ and $\delta_b$
at the two end points of the interval $A$ i.e. $x=a$ and $x=b$, respectively.

The entanglement entropy $S_A$ can be computed using the standard replica trick, i.e. by taking derivative
w.r.t. $n$ of the trace:
\ba
S_A=-\frac{\de}{\de n}\log\left[\mbox{Tr}_A \rho^n \right]\bigr|_{n=1}
=-\frac{\de}{\de n}\log \la \sigma_{n}(a)\bar{\sigma_{n}}(b)\lb_{\Sigma}\bigr|_{n=1}. \label{replica}
\ea
In this way, we can calculate the entanglement entropy from the two point function on a torus.

In particular, we are interested in the middle time region of the large subsystem limit (\ref{middle}).
In this region we can simplify the cut offs $\delta_a$ and $\delta_b$ in the torus coordinate as follows:
\ba
\delta_a\simeq \frac{2\s{r^2-l^2}}{t^2}\cdot\ep,\ \ \ \ \delta_b\simeq \frac{2\s{r^2-l^2}}{b^2}\cdot\ep.  \label{cutmid}
\ea

The endpoints of the subsystem $A$ given by $(x,\tau)=(a,0)$ and $(b,0)$ are mapped into the points $(\rho,\theta)=(0,\theta_a)$ and $(0,\theta_b)$,
where $\theta_{a,b}$ can be found from the conformation maps (\ref{cmapa}) and (\ref{cmapb}).  Here we have to be careful about the
analytical continuation of the Euclidean time $\tau$ to the real time by the identification $\tau=it$. For this we write the
`imaginary part' of $w$ and $-\bar{w}$ as $\theta$ and $\bar{\theta}$, respectively. Note that in general we have $\theta\neq\bar{\theta}$
except at $t=0$.
In the middle time region (\ref{middle}) with the large subsystem size limit, we find
\ba
&& \theta_a\simeq -\pi+\frac{2\s{r^2-l^2}}{t},\ \ \ \ \ \bar{\theta}_a\simeq \pi-\frac{2\s{r^2-l^2}}{t},\no
&& \theta_b\simeq \bar{\theta}_b\simeq \pi-\frac{2\s{r^2-l^2}}{b}.  \label{thev}
\ea

Therefore we can rewrite the two point function
of the twist operators on the original coordinate $(x,\tau)$ in terms of that on the torus coordinate $(\rho,\theta)$ 
by taking into account the transformation (\ref{cutmid}) of the cut off scale.
The further analysis of the two point functions on the torus depends on what kind of 2d CFTs we consider.

\section{Evolution of Entanglement Entropy in Holographic CFTs}

In the holographic CFTs, there is a sharp phase transition for the partition function on a torus. We have the following two phases:
\ba
&& (i)\ \mbox{High temperature phase}:\ \beta<2\pi\ \ \mbox{or, equally}\ \ \frac{l}{r}>\frac{1}{\cosh(\pi)} \no
&& (ii)\ \mbox{Low temperature phase}:\ \beta>2\pi\ \  \mbox{or, equally}\ \ \frac{l}{r}<\frac{1}{\cosh(\pi)} .
\ea

The phase $(i)$ is dual to the BTZ black hole with the temperature $T=1/\beta$. On the other hand, the phase $(ii)$ is dual to the thermal AdS geometry at the same temperature.
In both backgrounds, the spatial coordinate $\theta$ is compactified as $\theta\sim \theta+2\pi$.

\begin{figure}
  \centering
  \includegraphics[width=5cm]{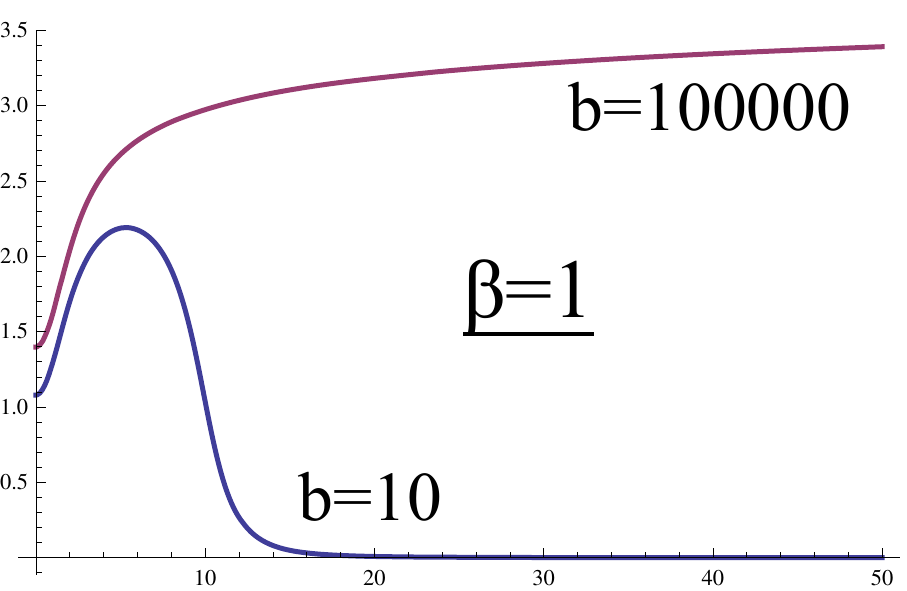}
 \includegraphics[width=5cm]{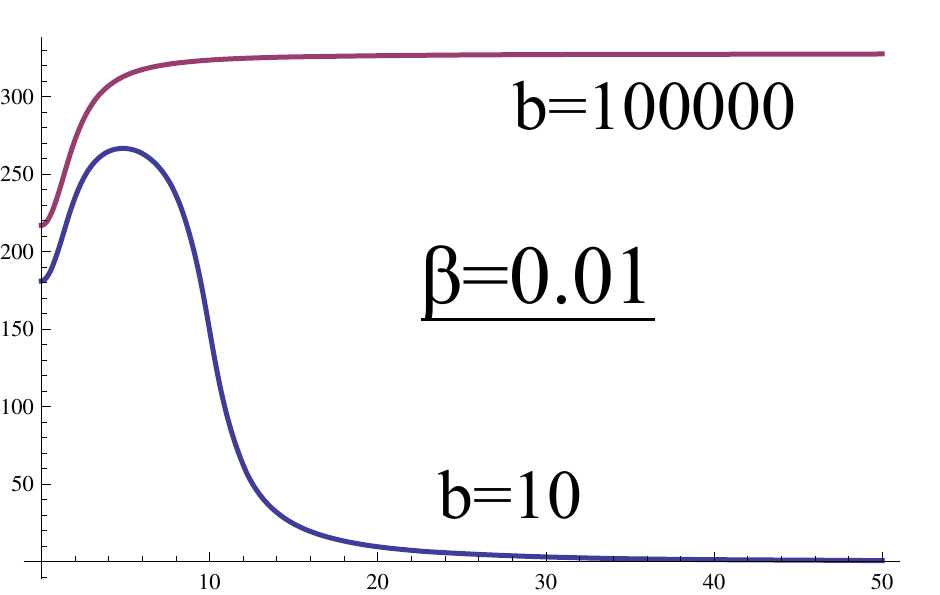}
 \includegraphics[width=5cm]{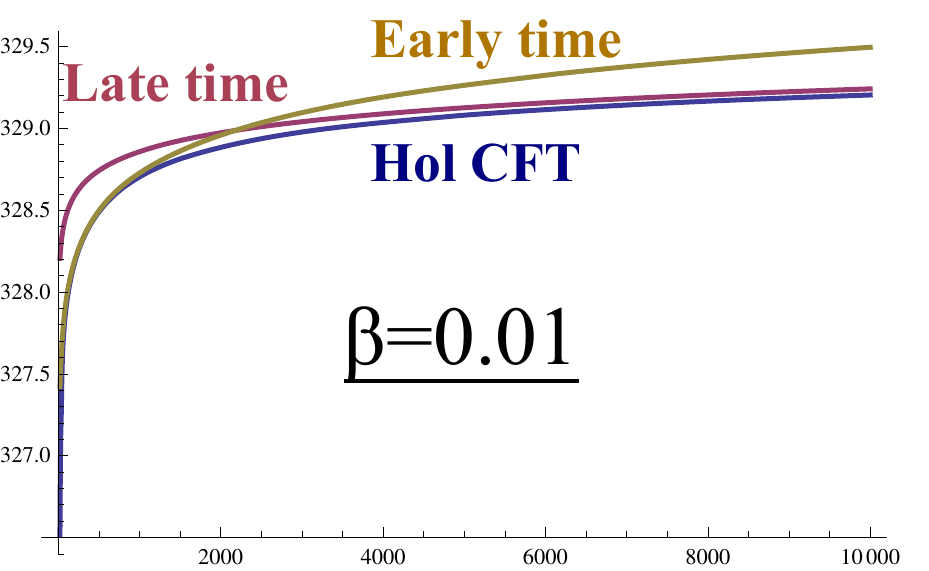}
 \caption{The plot of $\Delta S_A=S_A-S^{(0)}_A$ as a function of time $t$ in holographic CFTs. We set $c=1$ in all plots. The left and middle graph shows the plot for $\beta=1$ and $\beta=0.01$, respectively, both in the high temperature phase.
We took $a=1$, $r=2$ and $l=1$. The blue and red graph correspond to $b=10$ and $b=100000$.
The right picture shows the actual entanglement entropy (blue graph), the late time approximation (\ref{timh}) of the entanglement entropy (red graph), and the early time approximation (\ref{timhh}) of the entanglement entropy (yellow graph),
for  $\beta=0.01$, $a=1$, $b=100000$, $r=2$ and $l=1$. }
\label{Hentanglement entropyplot}
\end{figure}

\subsection{High Temperature Phase}

In the phase $(i)$ i.e. $\beta<2 \pi$, the entanglement entropy is computed by regarding the torus as an infinitely long cylinder extending into the
$\theta$ direction with the coordinate $\rho$ compactified as $\rho\sim \rho+\beta$.
Thus we find
\ba
S_A=\frac{c}{6}\log\left[\frac{\beta^2}{\pi^2\delta_a\delta_b}
\sinh\left(\frac{\pi(\theta_a-\theta_b)}{\beta}\right)\sinh\left(\frac{\pi(\bar{\theta}_a-\bar{\theta}_b)}{\beta}\right)\right]. \label{sah}
\ea
Let us first assume $\beta$ is $O(1)$.
For the time region (\ref{middle}), we can plug (\ref{thev}) into (\ref{sah}) and obtain the time evolution as follows:
\be
S_A\simeq \frac{c}{3}\log\left[\frac{\beta b t}{2 \pi \ep \s{r^2-l^2}}\right]
+\frac{c}{6}\log\sinh\left[\frac{2\pi\s{r^2-l^2}}{\beta t}\right]
+\frac{c}{6}\log\sinh\left[\frac{2\pi^2}{\beta}-\frac{2\pi\s{r^2-l^2}}{\beta t}\right].
\label{timev}
\ee

Thus we obtain (we used the relation (\ref{rel}))
\be
S_A\simeq \frac{c}{3}\log\frac{b}{\ep}+\frac{c}{6}\log\frac{t}{s}+\frac{c}{6}\log\left[\frac{\beta}{2\pi}\sinh\left(\f{2\pi^2}{\beta}\right)\right].
\label{timh}
\ee
Refer to the left picture in Fig.\ref{Hentanglement entropyplot} for a plot of $\Delta S_A$, namely the difference
between the entanglement entropy for our excited state and that for the ground state (\ref{vacentanglement entropy}).

If $\beta$ is very small $\beta\ll 1$ (i.e. high temperature limit), there is an additional intermediate behavior in the time region
\ba
s\ll t \ll \frac{s}{\beta},   \label{middlebeta}
\ea
where we find
\be
S_A\simeq \frac{c}{3}\log\frac{b}{\ep}+\frac{c}{3}\log\left(\frac{\beta t}{4\pi s}\right)+\frac{\pi^2 c}{3\beta}. \label{timhh}
\ee
However, for $t \gg \frac{s}{\beta}$ we find the previous behavior (\ref{timev}). Notice that the coefficient of $\log t$ is doubled in
(\ref{timhh}) compared with the previous one (\ref{timev}).
Refer to the middle and right picture in Fig.\ref{Hentanglement entropyplot} for a plot.

\subsection{Low Temperature Phase}
In the phase $(ii)$ i.e. $\beta>2\pi$, the entanglement entropy is computed by regarding the torus as an infinitely long cylinder extending in the $\rho$ direction with
the coordinate $\theta$  compactified as $\theta\sim \theta+2\pi$. Thus we obtain
\ba
S_A=\frac{c}{6}\log\left[\frac{4}{\delta_a\delta_b}
\sin\left(\frac{\theta_a-\theta_b}{2}\right)\sin\left(\frac{\bar{\theta}_a-\bar{\theta}_b}{2}\right)\right]. \label{sal}
\ea
By explicit calculations, we find that the entanglement entropy remains the same as that for the
ground state:
\be
S_A= \frac{c}{3}\log\frac{b-a}{\ep},
\label{timhhh}
\ee
at any time $t$.
Therefore there is no non-trivial time evolution in the low temperature phase within the large $c$ approximation.

\section{Evolution of Entanglement Entropy in Free Dirac Fermion CFT}

As another example, which is extremely different from the holographic CFTs, we would like to calculate the entanglement entropy $S(\rho_A)$ in the massless Dirac fermion CFT ($c=1$)
in the same setup. The entanglement entropy on a torus in this CFT was computed in \cite{ANT}.
In our present computation, we need to further perform the analytical continuation of the time
and convert the UV cut off $\delta_{a.b}$ on the torus into the cut off $\ep$ for the original $\Sigma$ space via (\ref{thev}).

The entanglement entropy in our setup is obtained from the result in \cite{ANT} as follows:
\ba
S_A&=&\frac{1}{6}\log\left[\frac{\beta^2}{\pi^2\delta_a\delta_b}\right]+g(\theta_b-\theta_a)+g(\bar{\theta}_b-\bar{\theta}_a),
\ea
where the function $g(y)$ is defined by
\ba
g(y)&=&\frac{1}{6}\log\sinh\left(\frac{\pi y}{\beta}\right)+\frac{1}{6}\sum_{m=1}^\infty \log\left[\frac{(1-e^{\frac{2\pi y}{\beta}}e^{-\frac{4\pi^2m}{\beta}})
(1-e^{-\frac{2\pi y}{\beta}}e^{-\frac{4\pi^2m}{\beta}})}{(1-e^{-\frac{4\pi^2m}{\beta}})(1-e^{-\frac{4\pi^2m}{\beta}})}\right] \no
&&+\sum_{p=1}^\infty \frac{(-1)^p}{p}\cdot
\frac{\left(\frac{\pi py}{\beta}\right)\coth\left(\frac{\pi py}{\beta}\right)-1}{\sinh\left(\frac{\pi py}{\beta}\right)}.
\ea

It is useful to examine the behaviors of $g(y)$. In the limit $y\to 0$ we have
\ba
&& g(y)\simeq \frac{1}{6}\log\frac{\pi y}{\beta}+O(y).
\ea
On the other hand, in the limit $y\to2\pi$ we find
\ba
&& g(y)\simeq \frac{1}{6}\log\left[\frac{\pi}{\beta}(2\pi-y)\right]+\frac{\pi^2}{3\beta}+\sum_{p=1}^\infty \frac{(-1)^p}{p}
\frac{\left(\frac{2\pi^2 p}{\beta}\right)\coth\left(\frac{2\pi^2 p}{\beta}\right)-1}{\sinh\left(\frac{2\pi^2 p}{\beta}\right)}+O(2\pi-y). \no
\ea

Below we consider the evolution of entanglement entropy in the middle time region with the large subsystem size limit (\ref{middle}).

\begin{figure}
  \centering
\includegraphics[width=5cm]{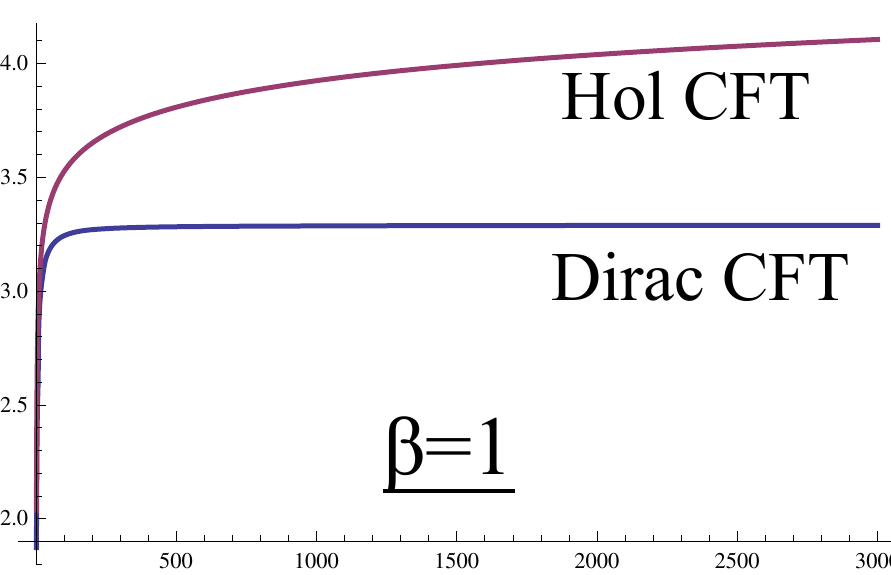}
  \includegraphics[width=5cm]{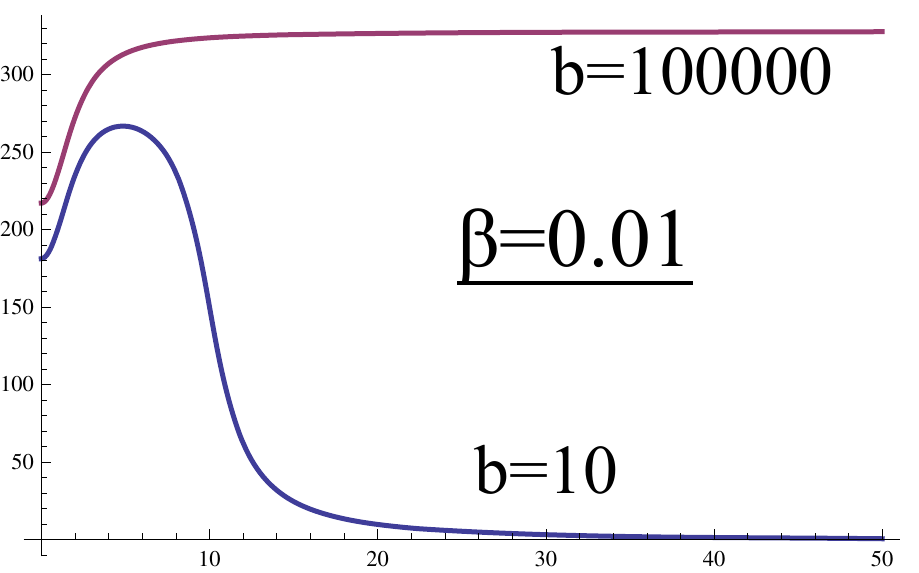}
 \includegraphics[width=5cm]{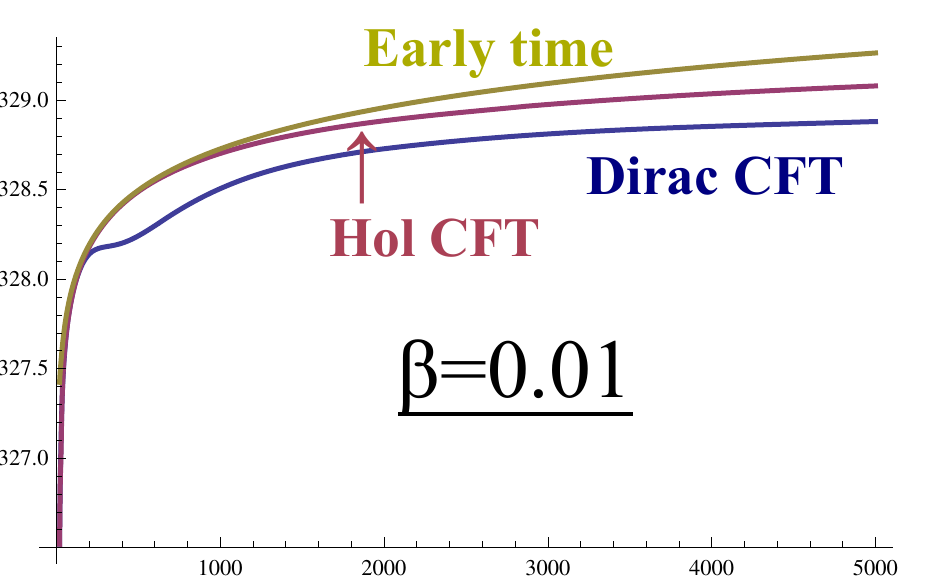}
 \caption{The plot of $\Delta S_A=S_A-S^{(0)}_A$ as a function of time $t$ for the Dirac fermion CFT.
The left picture is the comparison between the Dirac  fermion CFT result (blue graph) and
the holographic CFT result (red graph) at  $\beta=1$, $a=1$, $b=100000$, $r=2$ and $l=1$.
The middle graph shows the plot in the Dirac fermion CFT for $\beta=0.01$, $a=1$, $r=2$ and $l=1$. The blue and red graph correspond to the two difference values of $b$: $b=10$ and $b=100000$, respectively.
The right picture shows $\Delta S_A$ in the Dirac fermion CFT (Blue graph), the holographic CFT result (red graph), and the early time approximation (\ref{timhh}) entanglement entropy (yellow graph)
for  $\beta=0.01$, $a=1$, $b=100000$, $r=2$ and $l=1$.}
\label{Diplot}
\end{figure}

\subsection{Case 1: $\beta=O(1)$}

First we assume $\beta=O(1)$.  In this case for the time region (\ref{middle}), we find the behavior
\ba
S_A\simeq \frac{1}{3}\log\frac{b}{\ep}+\frac{1}{2}S_{th}(\beta),  \label{xa}
\ea
where $S_{th}(\beta)$ is the thermal entropy of the free Dirac fermion CFT at temperature $T=1/\beta$:
\be
S_{th}(\beta)=\frac{2\pi^2}{3\beta}+2\sum_{p=1}^\infty \frac{(-1)^p}{p}
\frac{\left(\frac{2\pi^2 p}{\beta}\right)\coth\left(\frac{2\pi^2 p}{\beta}\right)-1}{\sinh\left(\frac{2\pi^2 p}{\beta}\right)}.
\ee
Therefore, after the time $t=a$, when the signal of local operator quench started to affect $S_A$,
the entanglement entropy quickly increases by a half of the thermal entropy and stays the same.
Refer to the left picture in Fig.\ref{Diplot} for a comparison with the holographic result.

\subsection{Case 2: $\beta\ll 1$}

On the other hand, if we assume the very high temperature region $\beta\ll 1$, there is a time region:
\be
s\ll t \ll \frac{s}{\beta},  \label{timez}
\ee
where we have $\frac{2\pi}{\beta}(2\pi-y)\gg 1$. Therefore, in this time period,
we find the entanglement entropy behaves as
\ba
S_A\simeq \frac{1}{3}\log\frac{b}{\ep}+\frac{1}{3}\log\left(\frac{\beta t}{4\pi s}\right)+\frac{\pi^2}{3\beta}. \label{xb}
\ea
Note that this is identical to the holographic CFT result in the same time zone  as in (\ref{timhh}).

At the late time $t\gg \frac{s}{\beta}$, we find the previous behavior (\ref{xa}).
Note also at high temperature $\beta\ll 1$ we have $S_{th}(\beta)\simeq \frac{\pi^2}{3\beta}$.
Refer to the middle and right picture in Fig.\ref{Diplot} for explicit numerical plots.

\section{Bounds for Entanglement Entropy}

We would like to study the implications of the bounds (\ref{eeb}) for entanglement entropy
under the local operator quenches. We can rewrite the inequalities (\ref{eeb})  as follows:
\ba
 S(\rho)-H(p) \leq \sum_i p_i S(\rho_i)\leq S(\rho).  \label{bound}
\ea
This inequality is useful when $S(\rho)$ is calculable e.g. the case where $\rho$ is a canonical distribution or our universal local quench.
We can obtain the upper and lower bound for $S(\rho_i)$ i.e. the entropy for microscopic states or
the pure state local operator quench states.

To apply the entropy bound to our example of operator local quenches,
let us study the case where $p_i$ is given by the canonical distribution as in our universal local quenches:
\be
p_i=\frac{e^{-\beta \left(\Delta_i-\frac{c}{12}\right)}}{Z(\beta)}.
\ee
 The Shannon entropy is given by the thermal entropy as $H(p)=S_{th}(\beta)$.
 The microstates $\rho_i$ are now the reduced density matrices
$\rho^{(i)}_A$ defined in (\ref{lqop}). We identify the density matrix $\rho$ in (\ref{bound}) with the reduced matrix
 $\rho_A=\mbox{Tr}_B[\rho(\beta,s)]$, which is described as the mixed state
\ba
\rho_A=\sum_i  \frac{e^{-\beta\left(\Delta_i-\frac{c}{12}\right)}}{Z(\beta)}\cdot \rho^{(i)}_A.
\ea
This leads to the bounds from  (\ref{bound}):
\ba
S(\rho_A)-S_{th}(\beta)\leq \sum_i  \frac{e^{-\beta\left(\Delta_i-\frac{c}{12}\right)}}{Z(\beta)} S(\rho^{(i)}_A)\leq S(\rho_A).  \label{boundent}
\ea

If we write the density of states as $D(\Delta)$, the partition function looks like
\ba
Z(\beta)=\sum_{\Delta}D(\Delta)e^{-\beta\left(\Delta-\frac{c}{12}\right)}.
\ea
By imposing the modular invariance $Z(\beta)=Z(4\pi^2/\beta)$, we find in the limit $\beta\to 0$,
\be
Z(\beta)\sim e^{\frac{\pi^2c}{3\beta}}.
\ee
This leads to the Cardy formula as usual:
\ba
D(\Delta)\sim e^{2\pi\s{\frac{c}{3}\left(\Delta-\frac{c}{12}\right)}}.  \label{cardy}
\ea
For generic unitary CFTs, this Cardy formula (\ref{cardy}) can be applied when $\Delta \gg c$.
However, for holographic CFTs we can apply the Cardy formula (\ref{cardy}) when $\Delta>\frac{c}{6}$ \cite{Hart}.

In the high temperature phase (i.e. when we can apply the Cardy formula),
we can apply the saddle point approximation to an expectation value of a quantity $S(\Delta)$:
\ba
\frac{1}{Z(\beta)}\sum_{\Delta} D(\Delta)S(\Delta) e^{-\beta\left(\Delta-\frac{c}{12}\right)}
\simeq S(\Delta_\beta),
\ea
where we defined
\ba
\Delta_\beta=\frac{c}{12}+\frac{\pi^2 c}{3\beta^2},   \label{derelap}
\ea
or equally we have
\ba
\beta\simeq \frac{2\pi}{\s{\frac{12\Delta_\beta}{c}-1}}.  \label{hhh}
\ea

Thus at high temperature, the bound (\ref{bound}) can be rewritten as follows:
\ba
S^{ULOQ}_A(\beta)-S_{th}(\beta)\leq S^{LOQ}_A(\Delta_\beta) \leq S^{ULOQ}_A(\beta).  \label{boundd}
\ea

\subsection{Entanglement Entropy Bound for Holographic CFTs}

In the high temperature phase of holographic CFTs, by comparing (\ref{timh}) with (\ref{eeploq})
using (\ref{hhh}), we find that the latter inequality of (\ref{boundent}) is saturated i.e.
\ba
S^{ULOQ}_A(\beta)= S^{LOQ}_A(\Delta_\beta). \label{equr}
\ea
This is consistent with the indistinguishability  of the reduced density matrices $\rho^{(i)}_A$ as we trace out a larger part $B$.
This is analogous to the result in \cite{BO} for the examples of a finite temperature mixed state versus each micro states. 

On the other hand, in the low temperature phase, we have $S_{th}(\beta)=O(1)$ due to the confining nature. Therefore, the inequalities (\ref{bound}) are saturated, though we cannot apply (\ref{equr})
as the saddle point approximation breaks down at low temperature.
As we explained before, the entanglement entropy of the universal local operator quench state, becomes trivial in low temperature phase as in (\ref{timhhh}). This is consistent because in the discrete sum we find $e^{-\beta \Delta}\sim e^{-O(c)}$ is exponentially small
and thus we find $ \sum_i  \frac{e^{-\beta\left(\Delta_i-\frac{c}{12}\right)}}{Z(\beta)} S(\rho^{(i)}_A)$ is approximated by the
ground state entanglement entropy.

\subsection{Entanglement Entropy Bound for Dirac Fermion CFTs}
In the Dirac fermion CFT, we find $\Delta S^{ULOQ}_A(\beta)\leq \frac{1}{2}S_{th}(\beta)$ from the previous result (\ref{xa}).
Therefore the bound (\ref{boundd}) does not look tight. Nevertheless we obtain the upper bound
\be
\Delta S^{LOQ}_A(\Delta) \leq \frac{\pi^2}{3\beta}\simeq \frac{\pi}{6}\s{\frac{12\Delta}{c}-1}.
\ee
In particular, this shows there should be no logarithmic time growth at late time in the Dirac fermion CFT
as opposed to that in holographic CFTs.

\section{Bounds for R\'{e}nyi Entropy and Simple Examples}

The bounds for entanglement entropy (\ref{eeb}) can be generalized
to the R\'{e}nyi entropy. First,  we can generalize the lower bound (\ref{eeb}) by replacing the von-Neumann entropy with the R\'{e}nyi
entropy for $\alpha\in(0,1]$,
\begin{equation}
\sum_{i}p_{i}S_{\alpha}(\rho_{i})\leq S_{\alpha}(\rho)\ (0<\alpha\leq1),\label{eq:RenyiBoundsLower}
\end{equation}
where $S_{\alpha}(\rho)\equiv1/(1-\alpha)\log{\rm Tr}\rho^{\alpha}$. However, it is well known that the R\'{e}nyi entropy does
not satisfy the concavity $\sum_{i}p_{i}S_{\alpha}(\rho_{i})\leq S_{\alpha}(\rho)$ for $\alpha>1$. This bound is saturated if the microstates $\rho_i$ are indistinguishable (regardless to $\alpha$).

On the other hand, the upper bound by classical R\'{e}nyi entropy $H_\alpha(p)=1/(1-\alpha) \log (\sum_i p_i^\alpha)$ is always true for any $\alpha$ when $\rho_i$ are pure (refer to e.g. \cite{Bosyk}),
\begin{equation}
S_{\alpha}(\rho) \le \sum_{i}p_{i}S_{\alpha}(\rho_{i})+H_{\alpha}(p)\ =H_{\alpha}(p)
\label{eq:RenyiBoundsUpper}.
\end{equation}

In order to compute the R\'{e}nyi entropy $S_{n}(\rho_{A})$ for the universal
local operator quench state (\ref{mixL}) for arbitrary $n\in\mathbb{Z}$,
one needs to deal with higher genus partition functions. Here instead we deal
with a simpler example of mixed state which is given by the mixture
of two locally excited states
\begin{align}
\rho_{A}(t) & ={\rm Tr}_{B}[p\ket{O_{1}(t)}\bra{O_{1}(t)}+(1-p)\ket{O_{2}(t)}\bra{O_{2}(t)}]\\
 & \equiv p\rho_{1}(t)+(1-p)\rho_{2}(t),
\end{align}
where $0\leq p\leq1$. We will take $O_{1}=e^{i\eta\phi}$ and $O_{2}=e^{-i\eta\phi}$
with $h_{1}=\bar{h}_{1}=h_{2}=\bar{h}_{2}=\frac{\eta^{2}}{2}$ in
a free scalar field theory. The local operators are inserted at $X_{1}\equiv X_{3}\equiv ir_{1}$
and $X_{2}\equiv X_{4}\equiv-ir_{2}$ in the $X$ plane (we have used
$r=s$ for $l\to0$). 

\subsection{The second R\'{e}nyi entropy}

The second R\'{e}nyi entropy is given by 
\begin{equation}
S_{2}(\rho_{A})=-\log[p^{2}{\rm Tr}\rho_{1}^{2}+2p(1-p){\rm Tr}(\rho_{1}\rho_{2})+(1-p)^{2}{\rm Tr}\rho_{2}^{2}].
\end{equation}
We can compute these traces by using the following conformal map from
$X$-plane to $z$-plane (see \cite{HNTW}),
\begin{equation}
z^{2}=\frac{X-a}{X-b},
\end{equation}
and for the interacting term, we get 
\begin{align}
{\rm Tr}\rho_{1}\rho_{2} & =\frac{\braket{O_{1}^{\dagger}(X_{1},\bar{X_{1}})O_{1}(X_{2},\bar{X_{2}})O_{2}^{\dagger}(X_{3},\bar{X}_{3})O_{2}(X_{4},\bar{X_{4}})}_{(2)}}{\braket{O_{1}^{\dagger}(X_{1},\bar{X}_{1})O_{1}(X_{2},\bar{X}_{2})}\braket{O_{2}^{\dagger}(X_{3},\bar{X}_{3})O_{2}(X_{4},\bar{X}_{4})}}\frac{Z_{(2)}}{(Z_{(1)})^{2}} \nonumber \\ 
= & \prod_{k=1}^{4}\left(\frac{dz_{k}}{dX_{k}}\right)^{h_{k}}\left(\frac{d\bar{z}_{k}}{d\bar{X}_{k}}\right)^{\bar{h}_{k}}\frac{\braket{O_{1}^{\dagger}(z_{1},\bar{z}_{1})O_{1}(z_{2},\bar{z}_{2})O_{2}^{\dagger}(z_{3},\bar{z}_{3})O_{2}(z_{4},\bar{z}_{4})}}{\braket{O_{1}^{\dagger}(X_{1},\bar{X}_{1})O_{1}(X_{2},\bar{X}_{2})}\braket{O_{2}^{\dagger}(X_{3},\bar{X}_{3})O_{2}(X_{4},\bar{X}_{4})}}\frac{Z_{(2)}}{(Z_{(1)})^{2}}.
\end{align}
where 
\begin{align}
z_{1} & =-z_{3}=\sqrt{\frac{a-t-is}{b-t-is}},\ z_{2}=-z_{4}=\sqrt{\frac{a-t+is}{b-t+is}}.\\
\bar{z}_{1} & =-\bar{z}_{3}=\sqrt{\frac{a+t+is}{b+t+is}},\ \bar{z}_{2}=-\bar{z}_{4}=\sqrt{\frac{a+t-is}{b+t-is}}.
\end{align}

For the operators $O_{1}=e^{i\eta\phi}$ and $O_{2}=e^{-i\eta\phi}$, we find
\begin{align}
\frac{(Z_{(1)})^{2}}{Z_{(2)}}{\rm Tr}\rho_{1}\rho_{2} & =(1-z)^{4h}(1-\bar{z})^{4h},
\end{align}
where $z$ and $\bar{z}$ are the conformal cross-ratios
\begin{align}
z & =\frac{z_{12}z_{34}}{z_{13}z_{24}},\ \bar{z}=\frac{\bar{z}_{12}\bar{z}_{34}}{\bar{z}_{13}\bar{z}_{24}},
\end{align}
where $z_{ij}=z_{i}-z_{j}$. In the $s\to0$ limit, we have two regimes
depending on $t$ \cite{HNTW}
\begin{equation}
(z,\bar{z})\to\begin{cases}
(0,0) & t\notin(a,b)\\
(1,0) & t\in(a,b)
\end{cases}.
\end{equation}
Note that $\bar{z}$ is not the complex conjugate of $z$ because
of the analytic continuation. 

\begin{figure}
\begin{centering}
\includegraphics[scale=0.5]{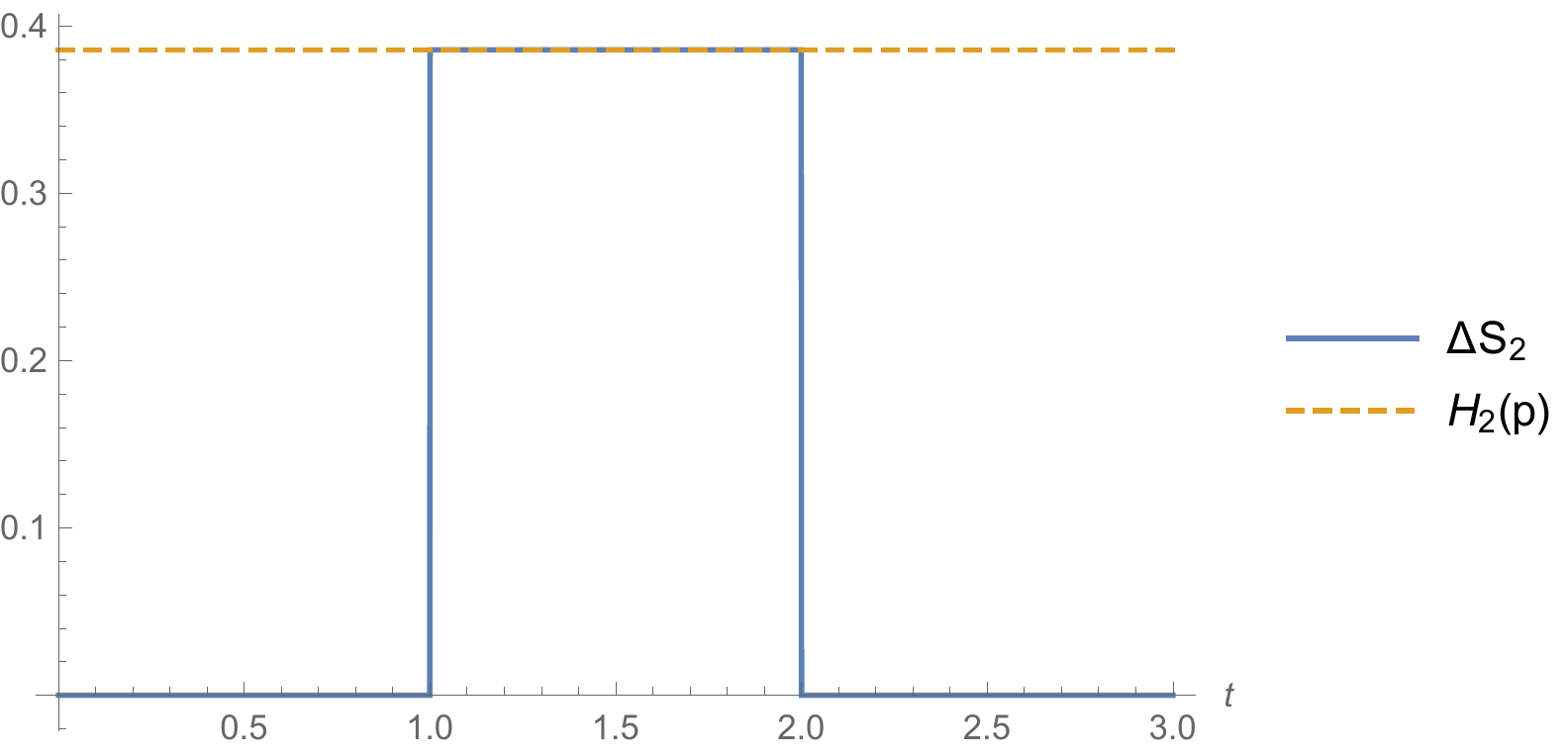}\caption{\label{fig:Evolution2ndRanyi}Evolution of the second R\'{e}nyi entropy
for $O_{1}=e^{i\eta\phi}$ and $O_{2}=e^{-i\eta\phi}$ for $p=0.2.$}
\par\end{centering}
\end{figure}

Similarly, we get (independent of $t$) 
\begin{align}
{\rm Tr}\rho_{1}^{2} & ={\rm Tr}\rho_{2}^{2}=\frac{Z_{(2)}}{(Z_{(1)})^{2}}.
\end{align}
Finally we get 
\begin{equation}
\Delta S_{2}(\rho_{A}(t))\equiv S_{2}(\rho_{A}(t))-S_{2}(\rho_{A}^{0})=\begin{cases}
0 & t\notin(a,b)\\
-\log\left(p^{2}+(1-p)^{2}\right) & t\in(a,b)
\end{cases}.  \label{wwwq}
\end{equation}
This is plotted in Fig.~\ref{fig:Evolution2ndRanyi}. Note that
one can understand the behavior in $a<t<b$ by the interpretation
of quasi-particle picture of the locally excited states (refer to \cite{Nozaki})
\be
e^{i\eta\phi}\ket{\Omega}\simeq\ket{0},\ \ \  \ e^{-i\eta\phi}\ket{\Omega}\simeq\ket{1} \label{qpp}.
\ee
 This leads to the interpretation 
\be
\rho_A\simeq p|0\lb\la 0|+(1-p)|1\lb\la 1|,
\ee
and this explains the non-trivial entropy in (\ref{wwwq}) for $a<t<b$. 
In this case the upper bound (\ref{eq:RenyiBoundsUpper}) is satisfied because $\rho_1$ and $\rho_2$ are
orthogonal to each other.

On the other hand, when $t>b$ or $t<a$ we find that the lower bound (\ref{eq:RenyiBoundsLower}) is satisfied (though it is not required) because all of the entropies are vanishing and $\rho_1$ and $\rho_2$ are indistinguishable.

\subsection{The $n$-th R\'{e}nyi entropy}

Moreover, one can generalize the analysis to the $n$-th R\'{e}nyi entropy in the $s\to0$ limit. In this case, we need to find 
\begin{equation}
{\rm Tr[}\rho_{i_{1}}\rho_{i_{2}}\cdots\rho_{i_{n}}]=\frac{\braket{O_{i_{1}}^{\dagger}(X_{1},\bar{X}_{1})O_{i_{1}}(X_{2},\bar{X}_{2})\cdots O_{i_{n}}^{\dagger}(X_{2n-1},\bar{X}_{2n-1})O_{i_{n}}(X_{2n},\bar{X}_{2n})}}{\prod_{j=1}^{n}O_{j}^{\dagger}(X_{2j-1},\bar{X}_{2j-1})O_{j}(X_{2j},\bar{X}_{2j})}\frac{Z_{(n)}}{Z_{(1)}^{n}},\label{eq:Trexpansion}
\end{equation}
for arbitrary $n\in\mathbb{Z}$ and arbitrary sets of the indices
$i_{k}=\pm1$ for $k\in\{1,2,\dots,n\}$ (we identify $O_{2}\equiv O_{-1}$). 

In the earlier time $t<a$ or the late time $b<t$, the $2n$-point function
appears in (\ref{eq:Trexpansion}) will factorize in the $s\to0$
limit as follows
\begin{equation}
\braket{O_{i_{1}}^{\dagger}(X_{1},\bar{X}_{1})O_{i_{1}}(X_{2},\bar{X}_{2})\cdots O_{i_{n}}^{\dagger}(X_{2n-1},\bar{X}_{2n-1})O_{i_{n}}(X_{2n},\bar{X}_{2n})}\sim\prod_{j=1}^{n}\braket{O_{i_{j}}^{\dagger}(X_{2j-1},\bar{X}_{2j-1})O_{i_{j}}(X_{2j},\bar{X}_{2j})}
\end{equation}
and thus we find a trivial behavior $\Delta S_{n}(\rho_{A}(t))=0$. 

On the other hand, for the middle time $a<t<b$, the $2n$-point function
is dominated by a non-trivial factorization 
\begin{equation}
\braket{O_{i_{1}}^{\dagger}(X_{1},\bar{X}_{1})O_{i_{1}}(X_{2},\bar{X}_{2})\cdots O_{i_{n}}^{\dagger}(X_{2n-1},\bar{X}_{2n-1})O_{i_{n}}(X_{2n},\bar{X}_{2n})}\sim\prod_{j=1}^{n}\braket{O_{i_{j}}^{\dagger}(X_{2j-1},\bar{X}_{2j-1})O_{i_{j}}(X_{2j-2},\bar{X}_{2j-2})}
\end{equation}
with the identification $X_{0}\equiv X_{n}$. By using
the conformal map 
\begin{equation}
z^{n}=\frac{X-a}{X-b},
\end{equation}
we can find (about the chiral part)
\begin{align}
{\rm Tr[}\rho_{i_{1}}\rho_{i_{2}}\cdots\rho_{i_{n}}] & \sim\prod_{j=1}^{n}\left(\frac{1}{n^{2n}}\frac{(z_{2j-1}^{n}-z_{2j}^{n})^{2}}{z_{2j-1}^{n-1}z_{2j}^{n-1}(z_{2j-1}-z_{2j-2})^{2(i_{2j-1}\cdot i_{2j-2})}}\right)^{h}\\
 & =\left(\varepsilon^{2(n-\sum_{j}(i_{2j-1}\cdot i_{2j-2}))}e^{\frac{2\pi i}{n}\sum_{j}(i_{2j-1}\cdot i_{2j-2})}\right)^{h},
\end{align}
where we have used 
\begin{equation}
z_{2k+1}=e^{\frac{2\pi i}{n}k}z_{1},\ z_{2k+2}=e^{\frac{2\pi i}{n}k}z_{2},
\end{equation}
and set set $z_{1}\equiv re^{-\frac{\pi i}{n}+i\varepsilon}$ and
$z_{2}=\bar{z}_{1}=re^{\frac{\pi i}{n}-i\varepsilon}$ with the infinitesimal
$\varepsilon$ related to $s$. Thus all the traces will
vanish in the $s\to0$ limit except the identical cases,
\begin{align}
{\rm Tr[}\rho_{i_{1}}\rho_{i_{2}}\cdots\rho_{i_{n}}] & =\begin{cases}
\frac{Z_{(n)}}{Z_{(1)}^{n}} & {\rm for\ }i_{1}=i_{2}=\cdots=i_{n}\\
0 & {\rm otherwise}
\end{cases}.
\end{align}

Therefore we finally obtain
\begin{equation}
\Delta S_{n}(\rho_{A}(t))=\begin{cases}
0 & t\notin(a,b)\\
\frac{1}{1-n}\log\left(p^{n}+(1-p)^{n}\right) & t\in(a,b)
\end{cases}.
\end{equation}
This is exactly what was expected from the quasi-particle picture. The inequalities (\ref{eq:RenyiBoundsLower}) and (\ref{eq:RenyiBoundsUpper}) are again saturated as in the second R\'{e}nyi entropy.

\section{More General Mixed States in Holographic CFTs}
Finally we use the inequality (\ref{eeb}) to constrain the evolution of entanglement entropy for mixed states
which are obtained by linear combinations of pure state local operator quenches with more general coefficients than the previous one (\ref{mixL}). We focus on holographic CFTs, to explore any possibilities of gravity duals for general mixed states.

\subsection{Setup}

We start with the general mixed states of the form,
\be
\rho^{mixed}=\sum_{i} p(\Delta_i)|\Psi_i(t)\lb\la\Psi(t)|,
\ee
where 
\ba \label{prob}
p(\Delta)=\frac{e^{-\gamma\, \Delta^{q}}}{Z}.  
\ea
Note that $\sum_i$ represents a sum of all possible states in a given CFT.
We assume $q>1/2$ for the convergence of the partition function $Z$. 
We also focus on the range $q\neq 1$ because $q=1$ is the case which we discussed in the previous sections.

Now to proceed, we first need to normalize $p(\Delta)$ in the following way,
\be \label{int}
\int_{0}^{\infty} p(\Delta)  D(\Delta) d\Delta=1,
\ee
where $D(\Delta)$ is the density of states. This determines the partition function $Z.$
In ideal holographic CFTs, we expect the following behaviors of the density of states:
\begin{align}
\begin{split}
D(\Delta) & \approx \mathcal{O}(1),  \quad\quad 0 < \Delta < \frac{c}{12},\\&
\approx e^{2\pi\s{\frac{c}{3}\left(\Delta-\frac{c}{12}\right)}},\quad \quad \frac{c}{12} <\Delta <\infty.
\end{split}
\end{align}
Though these requirements are more stronger than the one \cite{Hart}, the details do not affect our 
arguments in this section. Given this, to compute the Shanon entropy we will need to evaluate the following,
\begin{align}
\begin{split} \label{int1}
H(p)=-\sum_{i}p_i\log p_i=-\int_{0}^{\infty} p(\Delta) \log p(\Delta) D(\Delta) d\Delta.\\
\end{split}
\end{align}

Similarly, to compute the $\sum_{i} p_i S(\rho_i)$ we need to evaluate the  following,
\be \label{int2}
\sum_{i} p_i S(\rho_i)=\int_{0}^{\infty} D(\Delta) p(\Delta) S(\rho_\Delta) d\Delta. 
\ee
The expression for $S(\rho_i)$ is given in (\ref{entexp}). We will be evaluating the integrals appearing in  (\ref{int}), (\ref{int1}) and (\ref{int2}) by using saddle point approximation in the following section for various choice of $\gamma$. 

\subsection{Case 1: $\gamma=\mathcal{O}(1)$}

Let us first assume $\gamma$ remains order one when we take the large $c$ limit $c\to\infty$.
First notice that the saddle point $\Delta=\Delta_*$ of the integral of $\Delta$, if exists in the range $\Delta>\frac{c}{12}$, is specified by 
\ba
\pi\frac{\s{\frac{c}{3}}}{\s{\Delta_*-\frac{c}{12}}}-q\gamma \Delta_*^{q-1}=0.  \label{dafhre}
\ea
At this saddle point, we can estimate the contribution as follows
\ba
p(\Delta_*)D(\Delta_*)\sim \frac{1}{Z}e^{-\gamma \Delta_*^q +\frac{2\pi^2}{3q\gamma}c\Delta_*^{1-q}}.
\ea

Let us start with the case $q>1$. In this case, since $p(\Delta_*)D(\Delta_*)\sim e^{-\mathcal{O}(c^q)}/Z$ at the saddle point $\Delta_*>\frac{c}{12}$, we find that the dominant contribution comes from the region $\Delta \in [0, \frac{c}{12}]$. In this low energy region, due to the exponential damping factor of $p(\Delta)$, the integral is localized around at $\Delta=\mathcal{O}(1)$.  At the same time, for the integral (\ref{int1}), this clearly means that $H(p)=\mathcal{O}(1)$. Then the sum of entropy  (\ref{int2}) is also dominated by the contribution around $\Delta=\mathcal{O}(1)$. Since we know that the entanglement entropy $S_A$ for the state $|\Psi_i(t)\lb$ remains the same as that of the ground state in such a case, the same is true for $\sum_i p_i S(\rho_i)$. Note also that the inequalities (\ref{eeb}) get very tight because $H(p)=\mathcal{O}(1)\ll \sum_i p_i S(\rho_i)=\mathcal{O}(c)$. In summary, for $q>1$, we find the rather trivial result:
\ba
S(\rho^{mixed}_A)\simeq \sum_i p_i S(\rho_i)\simeq \frac{c}{3}\log\left(\frac{b-a}{\ep}\right).
\ea

For $1/2<q<1$ the dominant contribution comes from the the region $\Delta \in [ \frac{c}{12},\infty].$ 
The saddle point equation (\ref{dafhre}) shows that 
\ba
\Delta_*\simeq \left(\frac{\pi^2 }{3q^2\gamma^2}c\right)^{\frac{1}{2q-1}}\gg \mathcal{O}(c).
\ea
Then we get the estimations,
\be \label{ex}
H(p) \approx  \frac{2 \pi}{\s{3}}  \sqrt{c \left(\frac{c\,\pi^2}{3\gamma ^2 q^2}\right)^{\frac{1}{2 q-1}}} =\mathcal{O}(c^{\frac{q}{2q-1}}),
\ee
and
\be \label{ex1}
\sum_i p_i S(\rho_i) \approx   c^{\frac{q}{2 q-1}}\left(\pi^2/3\right) ^{\frac{q}{2 q-1}}  
\left(\gamma q  \right)^{-\frac{1}{2 q-1}} = \mathcal{O}( c^{\frac{q}{2 q-1}}),  
\ee
where we evaluated the last term in (\ref{eeploq}). 

In the above calculations, we have only kept the leading order term in the large $c$ limit. 
Now we see that expressions in (\ref{ex}) and (\ref{ex1}) are of the same magnitude, hence $S(\rho_P)$ cannot be approximated by $\sum_i p_i S(\rho_i).$ However, one might speculate 
$S(\rho^{mixed}_{A})=\mathcal{O}( c^{\frac{q}{2 q-1}})$ from the bounds (\ref{eeb}). Since $\frac{q}{2q-1}>1$, these contributions contradict with the expectations from the classical gravity description. Also we cannot  trust the formula (\ref{eeploq}) which was obtained from the standard large $c$ approximation in such a high energetic setup. In this way, for the range $1/2<q<1$, we cannot make any definite prediction for $S(\rho^{mixed}_A)$ from these arguments. 

\subsection{Case 2: $\gamma=\lambda\, c^{1-q}$} 

The previous difficulty in the case $1/2<q<1$ motivates us to scale the coefficient $\gamma$ in (\ref{prob}) 
appropriately in the large $c$ limit as follows
\ba
\gamma=\lambda\, c^{1-q},
\ea
such that the saddle point is $\Delta_*=\mathcal{O}(c)$ and the  contributions of various quantities are also $\mathcal{O}(c)$. The constant $\lambda$ in the above equation is taken to be $\mathcal{O}(1)$.

In this setup, we always find a dominant saddle point $\Delta=\Delta_*>\frac{c}{12}$ such that  $\Delta_*$ is $\mathcal{O}(c)$ as the solution to (\ref{dafhre}). When $q\lambda$ is very large or small, we find the behaviors
\ba
&&\mbox{When $q\lambda\ll 1$:}\ \ \ \ \Delta_*\simeq  
\left(\frac{\pi^2 }{3q^2\lambda^2}\right)^{\frac{1}{2q-1}}c,  \no
&&\mbox{When $q\lambda\gg 1$:}\ \ \ \ \Delta_*\simeq  \frac{c}{12}.
\ea

We can evaluate (\ref{int1}) and (\ref{int2}) as before and we find that both $H(p)$ and $\sum_i p_i S(\rho_i)$ are  $\mathcal{O}(c)$ in the large $c$ limit. In particular we obtain the behaviors:
\ba
&&\mbox{When $q\lambda\ll 1$:}\ \ \ H(q)\simeq 
 \frac{2 \pi^2}{3\lambda q}\cdot\left(\frac{\pi^2}{3q^2\lambda^2}\right)^{\frac{1-q}{2q-1}}\cdot c, \no
&&\mbox{When $q\lambda\gg 1$:}\ \ \ H(q)\simeq \frac{2\cdot 12^{q-1}\cdot \pi^2}{3\lambda q}\cdot c,
\ea
and
\ba
&&\mbox{When $q\lambda\ll 1$:}\ \ \ \sum_i p_i S(\rho_i)\simeq \frac{c}{3}\log\left(\frac{b-a}{\ep}\right)
+\frac{c}{6}\log\frac{t}{s}+\left(\pi^2/3\right) ^{\frac{q}{2 q-1}}  
\left(\lambda q  \right)^{-\frac{1}{2 q-1}} \cdot c, \no
&&\mbox{When $q\lambda\gg 1$:}\ \ \ \sum_i p_i S(\rho_i)\simeq \frac{c}{3}\log\left(\frac{b-a}{\ep}\right)
+\frac{c}{6}\log\frac{t}{s}+\frac{c}{6}\log\pi.
\ea

The inequalities  (\ref{eeb}) tells us that in the present case the entanglement entropy grows logarithmically:
\ba
S(\rho^{mixed}_A)\simeq \frac{c}{3}\log\left(\frac{b-a}{\ep}\right)
+\frac{c}{6}\log\frac{t}{s}+S_1,
\ea
where $S_1$ is a $\mathcal{O}(c)$ constant, though the inequalities are not enough tight to fix $S_1$ precisely. Also, the fact that the entanglement entropy is $\mathcal{O}(c)$ suggests a possibility that we can have a classical gravity dual of this mixed state.

\section{Conclusions}

In this paper, we introduced a new class of local quenches called universal local operator quenches. This is a mixed state counterpart of the standard local operator quenches \cite{Nozaki}, summing over infinitely many of the latter. These local quenches are parameterized by two parameters $\beta$ and $s$, where  $\beta$ plays a role of potential for the conformal dimensions of local operators we sum over, and $s$ is a regularization parameter of the local operators. After a conformal mapping, we found that it is described by a path-integral on a torus. Thus the entanglement entropy under a universal local operator quench can be computed as the two point functions of twist operators on a torus.  This allows us concrete computations of the time evolution of entanglement entropy. Note that our universal local operator quenches provide rare examples of mixed states for which we can analytically perform calculations using conformal field theoretic methods.

For explicit calculations of entanglement entropy, we focused on the two different CFTs in two dimensions: holographic CFTs and the massless Dirac fermion CFT. In holographic CFTs, our results of entanglement entropy under universal local operator quenches coincide with known results for the standard (pure state) local operator quenches, by relating $\beta$ in the former to the conformal dimension $\Delta$ of the local operator in the latter via (\ref{derelap}). Both of them grow logarithmically $\sim \frac{c}{6}\log t$ at late time under the time evolutions. We can conclude that this coincidence between the mixed state results and the pure state results is due to the chaotic nature of holographic CFTs. In other words, this is a consequence of ETH (eigenstate thermalization hypothesis) \cite{ETH,Kap} for two dimensional holographic CFTs. 

In the massless Dirac Fermion CFT, on the other hand, we found new behaviors from the time evolutions of entanglement entropy under universal local operator quenches. At late time, the growth of entanglement entropy approaches to a constant given by $\frac{1}{2}S_{th}(\beta)$, where $S_{th}$ is the thermal entropy when we consider the thermal partition function $Z(\beta)=\sum_i e^{-\beta(\Delta_i-c/12)}$.  This makes an intriguing contrast with the results for a local quench by a single primary operator, where the growth of entanglement entropy approaches to a constant given by the quantum dimension \cite{HNTW}.

We also found a common property for both the holographic and Dirac fermion CFT when $\beta$ is very small. In this case there is an intermediate time region $s\ll t\ll s/\beta$, where the entanglement entropy grows logarithmically $\sim \frac{c}{3}\log t$ with the doubled coefficient.

Moreover, in the latter parts of this paper, we discussed the implication of inequalities (\ref{eeb}).
In holographic CFTs, we found that one of inequalities is saturated, which is understood because 
we cannot distinguish various pure states due to the chaotic nature. In the Dirac fermion CFT, 
we obtain an upper bound from the inequality. This clearly proves that the entanglement entropy 
under pure state local operator quenches for any operator does not grow logarithmically as opposed 
to the entanglement entropy in holographic CFTs. We also discussed analogous inequalities for R\'{e}nyi 
entropy and analyzed a simple example of a free scalar CFT, where the saturations of inequalities occur.

Finally we explored what we can predict for the entanglement entropy in more general mixed states, 
by employing the inequalities (\ref{eeb}), when the probability distributions take the form 
$p_i\propto e^{-\gamma \Delta^q}$. Even though in general the analysis goes beyond our current knowledge, we manage to find a controllable example by fine-tuning $\gamma$ such that the resulting entanglement entropy is $O(c)$ and the bounds  (\ref{eeb}) provide meaningful predictions.

This paper initiates studies of entanglement entropy in general mixed states other than the standard finite temperature state. We expect such a research direction is also important for deeper understandings of AdS/CFT in order to work out what quantum states in holographic CFTs can have classical gravity duals. We can also think of various generalizations of our analysis in this paper, 
such as universal local operator quenches in higher dimensional  CFTs and multiple  local operator quenches. We would like to leave these for future problems.

\section*{Acknowledgements}

We are grateful to Pawel Caputa, Yoshifumi Nakata, Yuya Kusuki and Tomonori Ugajin for useful comments. TT is supported by the Simons Foundation through the ``It from Qubit'' collaboration and 
by World Premier International Research Center Initiative (WPI Initiative)  from the Japan Ministry of Education, Culture, Sports, Science and Technology (MEXT). AB and TT are supported by JSPS Grant-in-Aid for JSPS fellows 17F17023.  TT is supported by JSPS Grant-in-Aid for Scientific Research (A) No.16H02182 and by JSPS Grant-in-Aid for Challenging Research (Exploratory) 18K18766.
KU is supported by Grant-in-Aid for JSPS Fellows No.18J22888.

\end{document}